\documentclass[aps,twocolumn,showpacs,showkeys,floatfix,longbibliography,superscriptaddress]{revtex4-1}
\usepackage{rotating}
\usepackage{graphicx}
\usepackage{textcomp}
\usepackage{dcolumn}
\usepackage{amsmath}
\usepackage{amssymb}
\usepackage{epstopdf}
\usepackage{rotating}
\usepackage{color}
\usepackage{natbib}
\usepackage{xurl}
\setcitestyle{square}
\usepackage{textgreek}
\usepackage{multirow}
\usepackage{makecell}
\usepackage{setspace}
\usepackage{etoolbox}
\usepackage{lipsum}
\usepackage{booktabs}

\usepackage{array}
\newcolumntype{C}[1]{>{\centering\arraybackslash}p{#1}}

\usepackage{verbatim}
\usepackage[hypcap=true]{caption}
\usepackage{graphics}
\usepackage{tabularx}
\usepackage{hyperref}
\hypersetup{colorlinks=true,urlcolor=blue,citecolor=blue,linkcolor=blue}
\usepackage{soul}
\usepackage{enumitem}          % package for handling list formatting
\setlist{nosep}                 % Tightest spacing for lists. `noitemsep` is more relaxed

\usepackage{array}
\usepackage{ragged2e}

\newcolumntype{P}[1]{>{\RaggedRight\hspace{0pt}}p{#1}}

\newcommand{\fmtnum}[1]{%
  \ifnum\fpeval{#1 < 0.5} = 1
   \textcolor{red}{$#1$}%
   \else
   \textcolor{black}{$#1$}%
  \fi
}

\renewcommand{\arraystretch}{1.2} % Increase row height by 0.1

\begin{document}

\title{Mechanics Cognitive Diagnostic: Testing Fine-Grained Learning Objectives in Introductory Physics}

%not worrying about the order yet
\author{Vy Le}\affiliation{School of Education, Iowa State University, Ames, IA, 50011, USA}

\author{Jayson M. Nissen}\affiliation{Department of Physics, Montana State University, Bozeman, Montana 59715, USA} 

% \author{Xiuxiu Tang}\affiliation{College of Education, Purdue University, West Lafayette, IN, 47907, USA}

% \author{Yuxiao Zhang}\affiliation{College of Education, Purdue University, West Lafayette, IN, 47907, USA}

% \author{Amirreza Mehrabi}\affiliation{School of Engineering Education, Purdue University, West Lafayette, IN, 47907, USA}

\author{Jason W. Morphew}\affiliation{School of Engineering Education, Purdue University, West Lafayette, IN, 47907, USA}

\author{Hua Hua Chang}\affiliation{College of Education, Purdue University, West Lafayette, IN, 47907, USA}

\author{Ben Van Dusen}\affiliation{School of Education, Iowa State University, Ames, IA, 50011, USA}

\begin{abstract}
\textcolor{black}{Physics courses use research-based assessments (RBAs) such as the Force Concept Inventory (FCI), Force and Motion Conceptual Evaluation (FMCE), and Energy and Momentum Conceptual Survey (EMCS) to measure learning in introductory mechanics, but their fixed-length, pretest–posttest design makes them retrospective: posttest scores summarize completed instruction and arrive after a course ends. We are developing the Mechanics Cognitive Diagnostic (MCD), a cognitive diagnostic computerized adaptive test that reports students' mastery of fine-grained learning objectives (LOs) \textcolor{black}{throughout instruction}. Using evidence-centered design, we defined 14 LOs from introductory mechanics textbooks and AP Physics standards, mapped FCI, FMCE, and EMCS items onto them with a Q-matrix, and refined the mapping with the deterministic inputs, noisy ``and'' gate (DINA) model, using posttest responses from 24,394 students in 807 courses across 79 institutions through LASSO. The FCI and EMCS achieved good DINA model fit; the FMCE showed marginal fit. Classification accuracy for most LOs met or exceeded benchmarks for low-stakes formative assessment. RBA items, though not developed for LO-level diagnosis, support it reliably, giving the MCD a working 14-LO item bank built from \textcolor{black}{RBAs} that physics courses already use. As data accumulate, we can revise or retire weak LOs and items and add new items through online calibration without interrupting testing. We plan to expand the MCD to 35 LOs, two per week, to cover a typical introductory mechanics course.}

% energy content currently shows the weakest fit and smallest samples

\end{abstract}

\maketitle
\section{Introduction}
% New paper for Chain CAT

The Force Concept Inventory (FCI)~\citep{FCIhestenes1992force}, Force and Motion Conceptual Evaluation (FMCE)~\citep{FMCEthornton1998assessing}, and Energy and Momentum Conceptual Survey (EMCS)~\citep{EMCSsingh2003multiple} \textcolor{black}{are widely} used research-based assessments (RBAs) in introductory mechanics. These RBAs have shaped physics education research: they measure student content knowledge, document the effects of instructional reform, and enable comparisons across courses and institutions \citep{madsen2017resource, van2021online}. However, their fixed-length, pretest--posttest design positions them as retrospective or summative instruments. Pretest scores tell instructors how prepared students are at the start of the course, but they do not indicate which specific learning objectives
\textcolor{black}{(LOs; see Table~\ref{table:Def})} need attention as instruction unfolds. Posttest scores tell instructors how the semester went rather than what to spend more time on next week \citep{redish2003teaching}.

Effective formative assessment, by contrast, requires timely, actionable information about student thinking that instructors can use to adjust their teaching and students can use to direct their learning \citep{black1998assessment, black2004formative}. \textcolor{black}{\citet{leahy2005classroom} and \citet{wiliam2011embedded} identify two key aspects of effective formative assessment: sharing clear learning goals and providing feedback that moves learning forward.} Instructors administer the FCI, FMCE, and EMCS in a fixed-length, end-of-course format, and that format cannot support these practices. The end-of-course format withholds feedback until after the students whose responses generated it have finished the course \citep{black1998assessment, black2004formative}.

To address this gap, we developed the Mechanics Cognitive Diagnostic (MCD), the first cognitive diagnostic computerized adaptive testing (CD-CAT) assessment in physics education \citep{le2025applying}. We refer to that earlier version as MCD-v1 and to the LO-based version developed in this paper, which begins with 14 LOs, as MCD-v2. MCD-v1 mapped items from the FCI, FMCE, and EMCS onto four skills: apply vectors, conceptual relationships, algebra, and visualizations. Unlike a single fixed-length RBA, instructors could administer MCD-v1 adaptively throughout a semester, giving instructors and students skill-level mastery feedback during instruction rather than after it. MCD-v1 had a key shortcoming: its four skills were overly broad, combining conceptually distinct reasoning processes into a single attribute. For example, the ``visualization'' skill grouped graph reading in kinematics and force-diagram interpretation (tasks that draw on different knowledge and reasoning) so a low score on ``visualization'' did not indicate which specific aspect a student struggled with. CD model researchers have addressed this kind of over-broad attribute by subdividing content areas into more specific attributes (e.g., splitting forces into free-body diagrams and Newton's second law), but finer-grained attribute structures come at a cost: simulation studies show that increasing the number of attributes degrades item parameter recovery and classification accuracy \citep{de2010factors, templin2010diagnostic}, and most CD model applications therefore limit themselves to three to eight attributes \citep{kunina2012impact}. This tension between diagnostic specificity and classification accuracy has limited how far the field has subdivided attributes.

\begin{table*}[t!]
\centering
\caption{\justifying Definitions of terms.}
\label{table:Def}
\setlength{\tabcolsep}{5pt}
\renewcommand{\arraystretch}{1.3}
\begin{tabularx}{\textwidth}{>{\raggedright\arraybackslash}p{3.9cm}X}
\hline 
\hline 
Term & Definition\\ \hline %\smallskip
Learning objectives (LOs) & \textcolor{black}{In this study, LOs are the discrete knowledge components students must master to answer assessment items correctly, defined to align with the instructional goals and pacing that structure a physics course. The cognitive diagnostic literature typically calls these attributes or skills} \citep{chang2015psychometrics, helm2022cognitive, li2022use}. \\
Cognitive diagnostic (CD) assessment & \textcolor{black}{An assessment approach that classifies each student as having mastered or not mastered each of multiple LOs, rather than producing a single overall proficiency score} \citep{ravand2015cognitive, de2014cognitively}. \\ %\smallskip
Computerized adaptive testing (CAT) & A computer-based testing approach to select items sequentially based on a student’s previous responses to match item difficulty with the student’s estimated proficiency \citep{morphew2018using, chang2015psychometrics, weiss1982improving}. \\ %\smallskip
Evidence-centered design (ECD) & A framework for developing educational assessments that links claims about student knowledge to observable evidence and measurement models \citep{mislevy2003brief}. \\
Deterministic inputs, noisy ``and'' gate (DINA) model & A cognitive diagnostic model that assumes students must master all required LOs to answer an item correctly. The model represents each LO as either mastered or not mastered and does not allow one skill to compensate for another \citep{haertel1984application, junker2001cognitive, de2014cognitively, delaTorre2009}. \\ %\smallskip
Proficiency & \textcolor{black}{A continuous latent trait from item response theory representing} ``...the student’s general facility with answering the items correctly on the assessment under consideration'' \citep{stewart2023quantitative}. Higher proficiency increases the probability of answering assessment items correctly. Related terms used in the literature include ability, skill, latent trait, and omega. \\ %\smallskip
Q-matrix & A binary matrix that specifies the relationship between assessment items and the LOs required to answer them. Each row represents an item, and each column represents an LO. A value of 1 indicates that the item requires the LO to solve, whereas a value of 0 indicates that it does not. \\
Classification accuracy & \textcolor{black}{The degree to which a CD model's estimated mastered/not-mastered classification for each LO matches a student's true classification. Because true classifications are never directly observable, researchers estimate classification accuracy from the model's classification probabilities \citep{wang2015attribute}.} \\ %\smallskip
\hline
\hline
\end{tabularx}
\end{table*}

This study develops MCD-v2 and addresses these limitations by redefining the attribute structure in terms of LOs. We sized LOs to be finer-grained than weekly instructional units, at approximately two LOs per week, yielding 30--35 LOs across a semester \citep{beatty2013standards, richard2022implementing}. This grain size gives instructors meaningful diagnostic resolution while keeping the LOs interpretable and actionable. Building an item bank for that many LOs from scratch would take years. The FCI, FMCE, and EMCS contain items on much of introductory mechanics, and LASSO collects their responses at scale. This study tests these existing items as an item bank for a first set of 14 fine-grained LOs. We measure the coverage of the 14 LOs across the three assessments and the accuracy of a cognitive diagnostic model in recovering each student's mastery of each LO. The two research questions below formalize these aims.

\section{Research Questions}
We frame this study within the evidence-centered design (ECD) framework \citep{mislevy2003brief}. In ECD, the \textit{student models} specify the knowledge an assessment makes claims about, and the \textit{evidence models} specify the link between observed responses and those claims. In MCD-v2, the student models are the set of 14 LOs, and the evidence models are the Q-matrix and the deterministic inputs, noisy ``and'' gate (DINA) model connecting RBA items to those LOs (see Table~\ref{table:Def}). This study gathers validity evidence for the student models and asks two research questions.
\begin{enumerate}
    \item To what extent do the FCI, FMCE, and EMCS cover the 14 LOs defined for introductory mechanics? Specifically, how many LOs does each assessment measure, how many items measure each LO, and how many LOs does each item require?
    \item How well do DINA models fit each RBA's response data (two fit indices, RMSEA2 and SRMSR), and how accurately do they classify students' mastery of each LO?
\end{enumerate}
The first question establishes the coverage and structure of the student models across the three assessments. The second question tests the evidence models' recovery of that structure and identifies the LOs with strong and weak supporting evidence.

\section{Literature Review}
In this section, we review the RBAs used in introductory mechanics, followed by a discussion of formative assessment in physics education. We then review CD-CAT and prior applications of CAT in physics education.

\subsection{RBAs in introductory mechanics}

RBAs have played an important role in physics education research and in reforming teaching practices \citep{madsen2017resource}. Because RBAs use the same items across courses and institutions, they enable cross-course and cross-institutional comparisons. For example, researchers can benchmark outcomes against other courses and institutions \citep{hake1998interactive}, making them the primary evidence base for documenting the learning gains produced by active learning and evidence-based pedagogies in introductory physics \citep{von2016secondary}.

Online platforms have expanded access to RBAs. The Learning About STEM Student Outcomes (LASSO) platform \citep{LASSOv2} hosts RBAs, administers them to students online, scores the responses, and returns results to instructors and researchers \citep{van2021online}. PhysPort \citep{PhysPorttotalRBAs} offers RBAs and implementation guidance to instructors. LASSO matters for this study in two ways. It supplied the posttest responses we analyze, and it supports computerized adaptive testing \citep{LASSOv2}, so it can deliver MCD-v2 to students and support large-scale, fine-grained diagnostic assessment in introductory physics courses.

The FCI \citep{FCIhestenes1992force} is a 30-item multiple-choice assessment that probes students' conceptual understanding of force and motion, contrasting Newtonian reasoning with common informal beliefs about mechanics. Physics instructors use it more widely than any other mechanics RBA \citep{madsen2017resource}. The FMCE \citep{FMCEthornton1998assessing} covers content similar to the FCI, kinematics and Newton's laws, in 47 multiple-choice items. Several items share a physical scenario and ask students to select the response consistent with their reasoning about that scenario. The FMCE includes a small number of items on energy. The EMCS \citep{EMCSsingh2003multiple} is a 25-item multiple-choice assessment focused on energy and momentum, covering concepts such as kinetic and potential energy, work, and conservation principles that the FCI and FMCE do not address. Together, these three assessments cover the core topics of introductory mechanics.

\subsection{Formative Assessment in Physics Education}

Formative assessment refers to the ongoing process of gathering and using evidence of student learning to adjust instruction and support student progress \citep{black1998assessment}. \citet{black1998assessment} found that improving the quality of classroom formative assessment produces substantial gains in student achievement, with effect sizes from 0.4 to 0.7 across more than 250 studies, among the largest in educational research. Physics educators have used RBAs as formative tools to inform instruction, such as grouping students by prior knowledge or identifying topics that need reteaching \citep{madsen2016based}. However, their fixed-length, end-of-course design limits their formative utility: overall scores do not identify the specific LOs students struggle with, and the feedback arrives too late to benefit the students who generated it \citep{redish2003teaching}. \citet{yasuda2024chained} proposed shorter, more frequent assessments, but this alternative requires an item bank large enough to avoid overexposing students to the same items across repeated administrations.

Even a more frequently administered RBA would face a further barrier to fine-grained feedback: researchers have long debated the finer-grained constructs individual items measure \citep{huffman1995does, laverty2018analysis, stewart2018multidimensional, wells2020exploring}. For the FCI specifically, exploratory factor analysis failed to recover the FCI authors' own proposed sub-categories \citep{huffman1995does}, and subsequent studies have proposed differing factor structures, ranging from two to nine dimensions, without reaching consensus \citep{scott2012exploratory, stewart2018multidimensional, eaton2018confirmatory}. Without an agreed-upon mapping between items and constructs, an RBA score cannot reliably point to a specific topic, which is the actionable link formative assessment requires.

Organizing assessment around explicit LOs offers a direct path toward more usable formative feedback. Standards-based and competency-based grading frameworks in introductory physics structure instruction and assessment around defined LOs \citep{beatty2013standards, richard2022implementing}. One implementation raised course grades and lowered the rate of D and F grades and withdrawals, with the largest gains for women and first-generation students \citep{richard2022implementing}. When diagnostic feedback aligns with the same LOs that guide instruction, instructors can act on assessment results immediately, identifying which topics need additional support directly rather than inferring this from aggregate scores. Broad skill-based attributes, by contrast, group conceptually distinct reasoning processes, so a low score cannot identify which of those topics needs support.

\subsection{Cognitive diagnostic  computerized adaptive testing (CD-CAT)}

Computerized adaptive testing (CAT) provides a dynamic assessment approach grounded in item response theory (IRT), which models a probabilistic relationship between students' latent proficiency levels and their likelihood of correctly answering test items \citep{chang2015psychometrics}. In CAT, the algorithm updates the student’s proficiency estimate after each response and selects the next item using an information‑based criterion evaluated at the current estimate. In practice, this means selecting items matched to the current estimate of the student's proficiency \citep{chang2015psychometrics, morphew2018using}. This adaptive process keeps items neither too easy nor too difficult, yielding more precise proficiency estimates with fewer items than fixed paper‑and‑pencil tests.

A key advantage of CAT is maintaining measurement accuracy with reduced test length, often requiring fewer items than paper-and-pencil methods while incorporating content balancing to ensure the selected items cover the assessment's full range of topics \citep{csahin2017effects}.
Empirical evidence supports this efficiency: for instance, in an introductory physics course, CAT demonstrated moderate parallel-forms reliability and criterion-related validity, correlating with midterm exam scores while using fewer items than paper-and-pencil tests \citep{morphew2018using}.
CAT also enhances test security by drawing from large item banks to minimize overexposure \citep{chen2008controlling}.

Integrating cognitive diagnostic (CD) models with CAT, often termed cognitive diagnostic computerized adaptive testing (CD-CAT), further refines the assessment process by classifying students’ mastery status of multiple, distinct LOs that complement an overall proficiency score \citep{chang2015psychometrics, collares2022cognitive}. \textcolor{black}{CD models define LOs as fundamental cognitive units required to solve domain-specific items, and they classify each student on each LO as binary, mastered or not mastered, rather than as a continuous score \citep{helm2022cognitive, li2022use}.
Researchers have applied CD models across diverse domains to evaluate students' mastery of specific LOs, including: multiple strategic problem-solving behaviors in educational psychology \citep{zhang2021exploring}; computational thinking competencies in middle-school students \citep{li2022use}; introductory accounting LOs in vocational education \citep{helm2022cognitive}; and general domain-specific knowledge and clinical reasoning in healthcare professions education \citep{collares2022cognitive}.}

\begin{figure*}[t!]
\centering
    \includegraphics[scale=0.19]{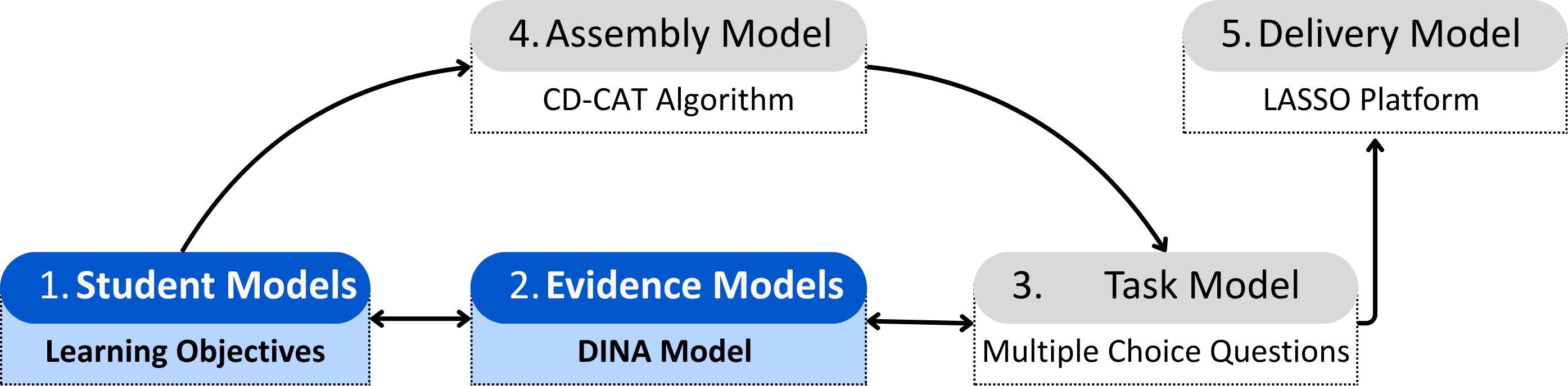}
    \caption{\justifying An evidence-centered design framework for creating the mechanics cognitive diagnostic (MCD). This paper focuses on the student models and evidence models (shown in black). The student models determine the learning objectives (LOs). The evidence models apply the DINA model to the multiple-choice questions (task model) that students answer to measure students' skills. Our CD-CAT algorithm will determine which items to ask students, who will take the assessment online through the LASSO platform.}
    \label{fig:framework}
\end{figure*}

\subsection{CAT in physics education}

While we do not know of any other research groups that have published on applying CD models in physics education, \textcolor{black}{researchers have begun employing CAT built on standard, single-scale IRT models to assess physics students' proficiency. For example, \citet{morphew2018using} found that CAT scores in an introductory calculus-based mechanics course correlated with midterm exam scores. Students who took a diagnostic CAT before an exam scored higher on that subsequent midterm than students who did not \citep{morphew2018using}. \citet{istiyono2018mapping} used a single-scale IRT-based CAT to place Indonesian senior high school students into overall physics problem-solving performance categories, rather than diagnosing mastery on multiple distinct constructs. Several studies focused on the FCI. \citet{yasuda2021analysis} reported that a two-parameter logistic model fit FCI-CAT data better than the alternative models they tested. \citet{yasuda2021optimizing} found that reducing test length to 15–19 items (50\%–63\% of original time) cost only 5\%–10\% in precision. \citet{yasuda2024chained} introduced chained CAT, a sequence of short adaptive tests spread across a semester. In their simulations, instructors could match the efficiency of traditional pre/post testing by giving students five items nine times per semester.}
Despite these advances in IRT-based CAT, none provided diagnostic feedback on the specific LOs or attributes students had mastered or not. 

\section{Theoretical Framework}

We drew on evidence-centered design (ECD) to guide the development of the MCD, following its use by researchers to design computer-based assessments \citep{mislevy2003brief,sheehan2007supporting}. In ECD, assessment design relies on three central premises \citep{mislevy2003brief}.

\color{black}{
\begin{enumerate}
    \item High-quality assessments require developers with both content and context expertise. We combined physics content knowledge with expertise in psychometrics, cognitive science, and physics education research to define LOs from mechanics textbooks and standards, and to build and validate a Q-matrix mapping FCI, FMCE, and EMCS items onto those LOs.
    
    \item Assessment developers rely on evidence-based reasoning to draw inferences about students’ understanding and misconceptions. We operationalized this through a Q-matrix, a table specifying the LOs each item requires. Our content-expert coding team built the initial Q-matrix, and we refined it using DINA model fit until it stabilized (see Section~\ref{sec:qmatrix} for details).
    
    \item Assessment developers must consider practical constraints, including available resources and testing conditions. The LASSO platform allowed us to administer multiple-choice items on web-enabled devices, reducing class time and instructor workload. We also sized LOs to roughly two per week of instruction. A quiz covering one week's LOs then needs about 12 items and 12 minutes, brief enough to fit inside regular class time.
\end{enumerate}}

ECD includes five models: student, evidence, task, assembly, and delivery models \textcolor{black}{(see Figure~\ref{fig:framework})}. This paper focuses on the student and evidence models. The student models identify key LOs that instructors aim to assess. The evidence models integrate evidence rules and measurement models to update information about each student’s performance: evidence rules define the way observable variables summarize item performance, and the measurement model converts item responses into a student's LO profile. Here, evidence rules use binary item scores, and the DINA model serves as the measurement model. The task model defines the actions and cognitive processes students engage in to produce evidence of learning. The assembly model integrates the student, evidence, and task models to establish the psychometric structure of the assessment. The delivery model employs the LASSO platform to administer assessments online.

The DINA model fit and classification-accuracy results below constitute \textit{validity evidence for the student models}: they test whether the LOs we defined \textit{a priori} from textbooks and AP test standards are measurable and separable in student responses. Strong fit and high classification accuracy for an LO indicate that the evidence supports that portion of the student models; weak fit or low accuracy indicate that the supporting evidence is limited, whether because of the items, the sample, or the measurement model. This framing also structures our treatment of limitations, which we cast as gaps in the current evidence for the student models, while noting where the evidence may point to revising an LO.

\section{Materials and Methods}

\subsection{RBA data collection and cleaning}
\label{sec:data}
We drew our dataset from the LASSO platform \citep{LASSOv2}. The data consisted of 69,449 responses from college students across three RBAs. We removed responses completed in fewer than five minutes. If a student completed the same assessment more than once, we kept their first attempt on the most recent posttest administration. After cleaning, the dataset contained 24,394 posttest responses from 315 algebra-based, 460 calculus-based, and 32 other physics courses across 79 institutions. Of these responses, 15,371 were from the FCI, 7,033 from the FMCE, and 1,990 from the EMCS. We excluded FCI item 29 from all analyses because of its poor psychometric performance, consistent with prior FCI factor analyses that dropped it \citep{scott2012exploratory, eaton2018confirmatory}.

\begin{table*}[t!]
\centering
\caption{\justifying Definition of the LOs in the FCI, FMCE, and EMCS assessments.}
\label{tab:LOs}
\begin{tabularx}{\textwidth}{p{4.0cm}X}
\hline 
\hline 
Learning Objectives & Definition \\ 
\hline
1. Vectors & Manipulate vector quantities in one or more dimensions, including changes in direction or sign. \\
2. 1D Kinematics & Analyze motion in one dimension using average or instantaneous velocity, speed, displacement, position, or acceleration. \\
3. Free Fall & Apply kinematic relationships to determine position, velocity, and/or acceleration for objects in free fall. \\
4. 2D Kinematics & Analyze motion in two dimensions, including projectile motion, using velocity, displacement, position, or acceleration. \\
5. Free-Body Diagrams & Identify the type and direction of the forces acting on an object. \\
6. Newton’s Second Law &  Apply Newton's first and second laws of motion to find mass, force, or acceleration in linear systems. \\
7. Newton’s Third Law & Apply Newton’s third law to analyze interactions and compare the direction and magnitude of forces between objects. \\
8. Kinetic Energy & Use the kinetic-energy relationship to determine an object’s kinetic energy, mass, or velocity. \\
9. Potential Energy & Determine gravitational or elastic potential energy and relate changes in potential energy to work done on a system. \\
10. Work & Apply the definition of work to determine force, displacement, the work done by a force, or relate work to changes in kinetic and potential energy. \\
11. Conserve Energy & Apply conservation of mechanical energy to calculate properties of simple systems. \\
12. Linear Momentum & Apply the definition of momentum to determine the momentum of objects or systems. \\
13. Impulse & Apply the impulse–momentum theorem to analyze interactions and solve collision problems. \\
14. Conserve Momentum & Determine whether a system is isolated and apply momentum conservation to solve collision or interaction problems. \\
\hline
\hline 
\end{tabularx}
\end{table*}

\subsection{Qualitative Data Analysis}
\label{Quali_Analysis}
We developed a list of LOs based on OpenStax physics textbooks and AP test standards. The OpenStax texts are peer-reviewed, \textcolor{black}{openly licensed}, and widely adopted in college-level introductory physics courses, ensuring alignment with rigorous university expectations \citep{openstaxphysv1,collegeAP2023}. \textcolor{black}{Each chapter also includes a detailed set of learning objectives, which served as our starting point}. The AP Physics 1 \citep{APphysics1} and AP Physics C - Mechanics \citep{APmechanics} represent the recognized learning outcomes that high school students must master to earn college credit, offering an exam-validated scope and sequence for introductory mechanics. \textcolor{black}{Using these sources with a goal of approximately two LOs a week}, we identified an initial set of LOs that span core topics such as vectors, kinematic motion, Newton’s laws, energy, and momentum (see Table~\ref{tab:LOs}).

Using this set of LOs, \textcolor{black}{we coded each assessment item for the LOs a student would need to master to answer it correctly}. Our coding team included three researchers with backgrounds in physics and physics education. At least two team members independently coded each item. The three coders then reviewed the results together and resolved any discrepancies through discussion. We then ran the quantitative analyses on the revised codes. This iterative process continued until a consensus was reached among all team members.

\subsection{Quantitative Data Analysis}

\subsubsection{Deterministic Inputs, Noisy ‘And’ Gate (DINA)}

MCD-v2 employs the DINA model, one of the most commonly used CD models \citep{de2014cognitively, junker2001cognitive}. In MCD-v2, each item requires students to master specific LOs to answer correctly. The DINA model encodes the relationship between items and required LOs through a Q-matrix, a binary (i.e., 0 or 1) matrix where rows represent items and columns represent LOs \citep{tatsuoka2012architecture}. An entry \( q_{jk} = 1 \) indicates that the \( j \)-th item requires the \( k \)-th LO, while \( q_{jk} = 0 \) indicates that it does not.

\textcolor{black}{The DINA model assumes that a student must have mastered all LOs required for a given item to answer it correctly; if the student lacks even one of those LOs, the model predicts an incorrect response \citep{delaTorre2009}}. The model uses the \textit{slipping} and \textit{guessing} parameters to account for real-world inconsistencies. The \textit{slipping} parameter represents the probability that students who have mastered all the required LOs may still produce incorrect answers, whereas the \textit{guessing} parameter represents the probability that students who lack some of the required LOs may answer correctly \citep{delaTorre2009}. We report the estimated DINA slipping and guessing parameters for every item in the Appendix (Tables~\ref{tab:itempars_fci}--\ref{tab:itempars_emcs}). 

We calibrated a three-parameter logistic (3PL) IRT model \citep{birnbaum1968some} for each RBA using the \texttt{mirt} package in R \citep{chalmers2012mirt}. We report its discrimination, difficulty, and pseudo-guessing parameters in the same tables. This follows \citet{le2025applying} and gives future CD-CAT work a set of item statistics to build on. The pseudo-guessing parameter flags items where low-proficiency students succeed at high rates, which helps interpret classification accuracy and model fit. The DINA model remains the measurement model for the MCD. We report the IRT parameters only to support replication and item-quality checks.

To evaluate DINA model fit, we used RMSEA2 and SRMSR from the G-DINA package in R \citep{ma2020gdina}. RMSEA2 is the root mean square error of approximation based on the M2 statistic, a limited-information fit statistic computed from item pairs \citep{maydeu2014assessing}. Values at or below 0.05 indicate good fit \citep{hooper2008evaluating, hu1999cutoff}. SRMSR is the standardized root mean square residual, the average discrepancy between observed and model-implied item correlations. Values at or below 0.07 indicate a well-fitting model \citep{maydeu2014assessing, hu1999cutoff}. \textcolor{black}{We used LO-level classification accuracy to evaluate the reliability and validity of the CD assessment. Following benchmarks in the CD model literature for low-stakes formative assessments, values $\geq 0.9$ indicate good, and values $\geq 0.8$ indicate acceptable accuracy \citep{tan2023tutorial, liang2023mental, paulsen2020examining}.}

\begin{table}[t!]
    \centering
    \caption{\justifying Q-matrix modifications and adoption rates.}
    \label{tab:Q_Val}
    \begin{tabular}{lcccc}
        \hline
        \hline
        & FCI & FMCE & EMCS & Overall \\
        \hline
        Total Items        & 29  & 47  & 25  & 101 \\
        Possible Changes   & 203 & 376 & 175 & 754 \\
        Suggested Changes  & 23  & 51  & 30  & 104 \\
        Adopted Changes    & 3   & 7   & 10  & 20 \\
        Adoption Rate      & 13\% & 14\% & 33\% & 19\% \\
        Change Rate        & 1.5\% & 1.9\% & 5.7\% & 2.7\% \\
        \hline
        \hline
    \end{tabular}
\end{table}

\begin{figure*}[t!]
\centering
    \includegraphics[scale=0.8]{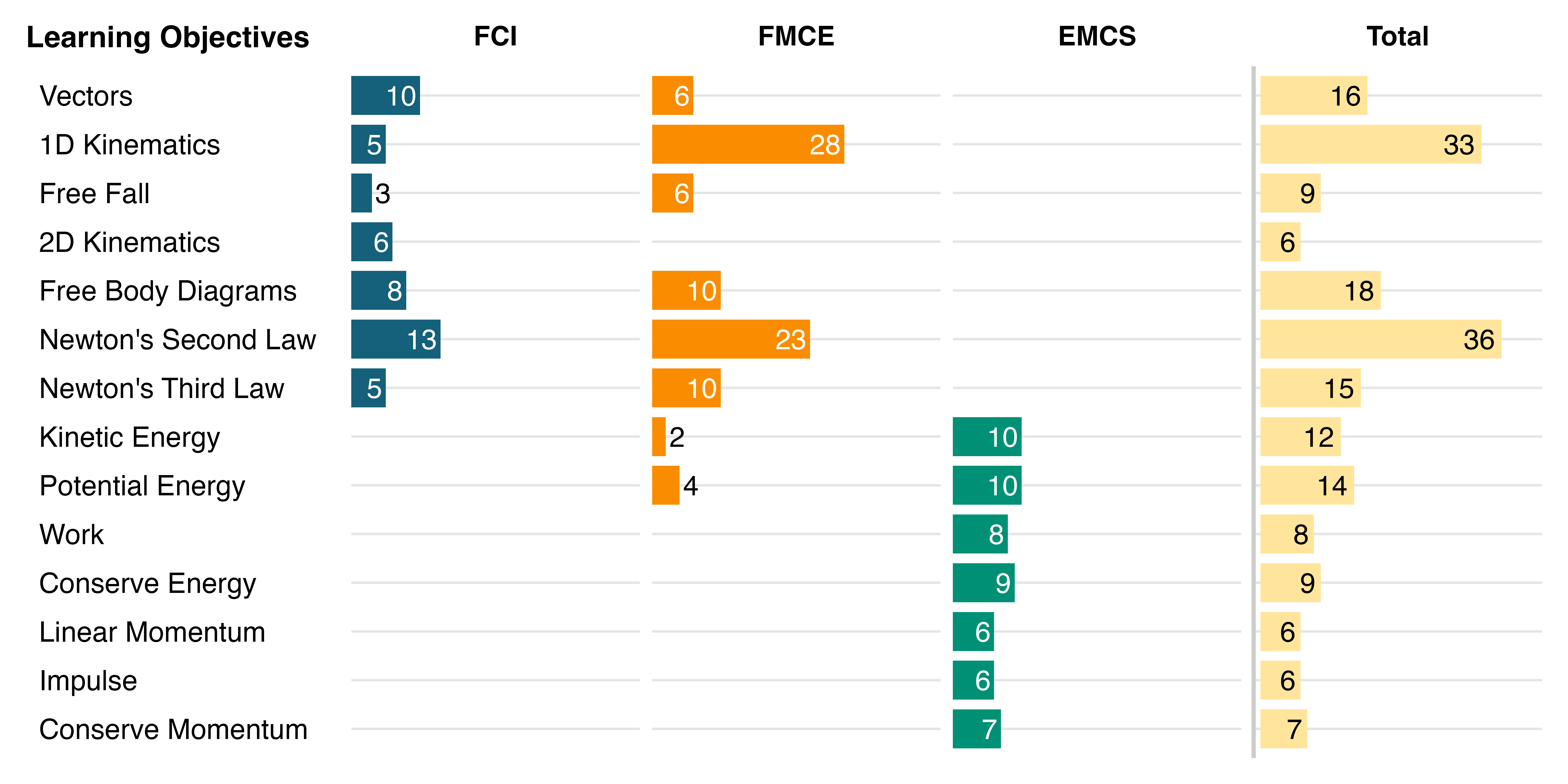}   
    \caption{\justifying Distribution of items across RBAs for all LOs.}
    \label{fig:FCI_FMCE_EMCS}
\end{figure*}

\subsubsection{Q-Matrix validation}
\label{sec:qmatrix}
\textcolor{black}{Because the Q-matrix specifies the model structure in CD assessments, its appropriateness directly determines model–data fit}. Misspecifications in the Q-matrix can lead to poor model fit and may produce inaccurate LO diagnoses for students. Therefore, researchers must validate the Q-matrix before conducting any diagnostic analysis. In this study, content experts constructed the Q-matrices used for the DINA analyses of each RBA, as described in Section~\ref{Quali_Analysis}. In the validation step described below, we examined each Q-matrix within the DINA framework to identify possible misspecifications. 

We analyzed students’ posttest responses for each of the three RBAs using separate DINA models. \textcolor{black}{The DINA model outputs suggested modifications to the Q-matrix, which the coders reviewed for consistency with the LO definitions. The team revised the Q-matrix when all three coders reached consensus on a suggested change.}

Table~\ref{tab:Q_Val} summarizes the frequency of data-driven modifications suggested by DINA, the number adopted by the coders, and the corresponding adoption rates for each RBA. The FCI, for example, had 23 proposed changes out of 203 possible changes (29 items each with seven possible LOs). The coders adopted three of these suggestions, resulting in an adoption rate of 13\%. Across all three RBAs, the analysis identified only 14\% of the codings (104 of 754) for re-examination. The coders reviewed all 104 suggestions and adopted 20 changes, resulting in an overall adoption rate of 19\%. 
\textcolor{black}{In total, this validation step changed only 2.7\% of all possible item-LO codings (20 of 754). This low overall change rate indicates that the initial expert coding closely matched the structure the DINA model recovered from student response data, and it supports the reliability of the final Q-matrix used in the subsequent CD modeling analyses reported below. Table~\ref{tab:topics} (see the Appendix) shows the final coding for each RBA item across the LOs.}

One adopted change illustrates how DINA can help identify ambiguities in expert coding. 
FCI item 1 asks students to compare the times required for two balls of different masses to fall the same distance. 
The initial coding mapped the item to Free Fall because students can obtain the correct answer using free-fall kinematics. 
DINA identified Newton’s Second Law as an additional required LO. Upon review, the coders reached consensus that distinguishing the correct answer from the expectation that gravity pulls the heavier ball harder and that it therefore falls faster requires students to recognize that the gravitational force on each ball is proportional to its mass and that the resulting acceleration is independent of mass. 
The coders adopted the change by consensus. 
This example illustrates the complementary roles of expert judgment and model-based evidence. 
DINA does not determine the Q-matrix, but it can identify item–LO relationships that warrant expert re-examination.

\section{Findings}

This section addresses the two research questions. We first present which of the 14 LOs the items on the three assessments measured and the number of LOs each item measured (RQ1), with the full item-LO coding listed in Table~\ref{tab:topics} in the Appendix. We then examine the model fit and classification accuracy (RQ2).

\subsection{Distribution of LOs Across RBAs}

Of the 14 LOs, the final Q-matrices mapped FCI items to 7 LOs, FMCE items to 8 LOs, and EMCS items to 7 LOs. Figure~\ref{fig:FCI_FMCE_EMCS} shows the distribution of items across LOs for each RBA. Across all three RBAs, Newton’s Second Law (36 items) and 1D Kinematics (33 items) had the greatest coverage, followed by Free-Body Diagrams (18), Vectors (16), and Newton’s Third Law (15). Energy- and momentum-related LOs were represented by 6–14 items each, largely drawn from the EMCS. The three RBAs covered a broad range of mechanics LOs, with kinematics and forces concentrated in FCI and FMCE, and energy and momentum concentrated in EMCS.

\begin{table}[t!]
\centering
\caption{\justifying Distribution of items across the number (\%) of learning objectives they assess.}
\label{tab:lo_count}
\setlength{\tabcolsep}{3pt}
\begin{tabular*}{0.48\textwidth}{c c c c c}
\hline 
\hline
\multirow{2}{*}{Number of LOs} 
 & \multicolumn{4}{c}{Assessment} \\
\cline{2-5}
 & FCI & FMCE & EMCS & Total \\
\hline
1 & 11 (38\%) & 18 (38\%) & 9 (36\%)  & 38 (38\%) \\
2 & 15 (52\%) & 16 (34\%) & 8 (32\%)  & 39 (39\%) \\
3 & 3 (10\%)  & 13 (28\%) & 4 (16\%)  & 20 (20\%) \\
4 & -         & -         & 2 (8\%)   & 2 (2\%)   \\
5 & -         & -         & 1 (4\%)   & 1 (1\%)   \\
6 & -         & -         & 1 (4\%)   & 1 (1\%)   \\
\hline
\hline
\end{tabular*}

\end{table}

Table~\ref{tab:lo_count} summarizes the number of LOs assessed by each item across the three RBAs. Most items measured one or two LOs. Across all assessments, 38 items (38\%) assessed a single LO and another 39 items (39\%) assessed two LOs. Twenty items (20\%) assessed three LOs, while only four items assessed four or more LOs.

\begin{figure*}[t!]
\centering
    \includegraphics[scale=0.8]{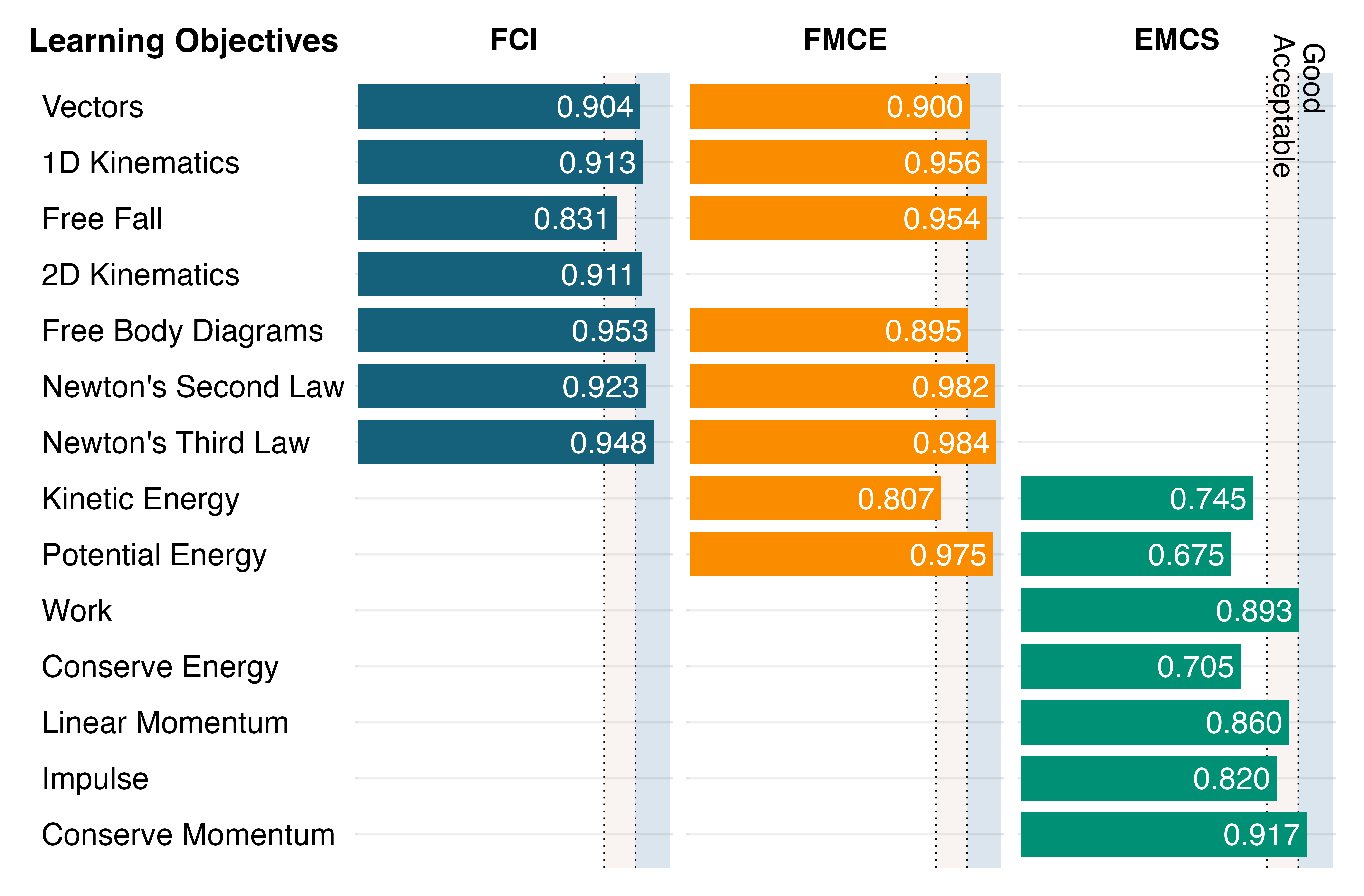}    
    \caption{\justifying Classification accuracy across LOs and RBAs. \textcolor{black}{The pink shaded region indicates acceptable classification accuracy (0.80--0.90) and the black shaded region indicates good classification accuracy ($\geq 0.90$).}}
    \label{fig:CA}
\end{figure*}

\textit{FCI} -- FCI items focused on kinematics and forces. It included items on Vectors (10), 1D Kinematics (5), Free Fall (3), 2D Kinematics (6), Free-Body Diagrams (8), Newton’s Second Law (13), and Newton’s Third Law (5). FCI did not include items on energy or momentum LOs. The FCI included items that targeted one or two LOs. Eleven items (38\%) assessed a single LO and 15 items (52\%) assessed two LOs. Only three items (10\%) assessed three LOs, and none assessed more than three.

\textit{FMCE} -- FMCE measured kinematics and forces, with strong coverage of 1D Kinematics (28) and Newton’s Second Law (23). It included additional items on Vectors (6), Free Fall (6), Free-Body Diagrams (10), and Newton’s Third Law (10), and had limited coverage of energy through Kinetic Energy (2) and Potential Energy (4). The FMCE showed a similar pattern with FCI in the distribution of items across LOs, but included more items that addressed three LOs. Eighteen items (38\%) assessed one LO and 16 items (34\%) assessed two LOs. Thirteen items (28\%) assessed three LOs, and none assessed more than three.

\textit{EMCS} -- EMCS items focused on energy and momentum. It included items on Kinetic Energy (10), Potential Energy (10), Work (8), Conserve Energy (9), Linear Momentum (6), Impulse (6), and Conserve Momentum (7) LOs and did not include items for kinematics or forces. The EMCS included items that addressed a broader range of LOs per item. Nine items (36\%) assessed one LO and eight items (32\%) assessed two LOs. Four items (16\%) assessed three LOs, while four items assessed four or more LOs. Specifically, two items (8\%) assessed four LOs, one item (4\%) assessed five LOs, and one item (4\%) assessed six LOs.

\begin{table}[t!]
\centering
\caption{\justifying Model fit values across three assessments. Values of RMSEA2 $\leq 0.05$, SRMSR $\leq 0.07$ indicate good model fit.}
\label{tab:modfit}
\begin{tabular*}{0.9\linewidth}{l@{\extracolsep{\fill}}l c c c} 
\hline
\hline
& FCI & \centering FMCE &  \centering EMCS\arraybackslash \\ 
\hline
RMSEA2  & 0.033 & 0.065 &  0.022 \\
SRMSR  & 0.055 & 0.087  & 0.037 \\
\hline
\hline
\end{tabular*}
\end{table}

\subsection{DINA model fit}
We fit the DINA model with the refined Q-matrix to the response data separately for each RBA (Table~\ref{tab:modfit}). The FCI yielded RMSEA2 = 0.033 and SRMSR = 0.055. The EMCS yielded RMSEA2 = 0.022 and SRMSR = 0.037. Both assessments met the criteria for good model fit. \textcolor{black}{The FMCE yielded RMSEA2 = 0.065 and SRMSR = 0.087, falling outside the criteria (RMSEA2 $\leq 0.05$, and SRMSR $\leq 0.07$). These results indicate that the DINA model does not reproduce the FMCE response data as well as it does for the FCI and EMCS. We take up the practical significance and possible sources of this misfit in Section~\ref{sec:fit}.}
 
\subsection{DINA model classification accuracy}

Figure~\ref{fig:CA} presents the classification accuracy \citep{wang2015attribute} for each LO across the three RBAs. \textcolor{black}{Combining 14 LOs with 3 RBAs gives 42 possible LO-assessment combinations, but each RBA covers only a subset of the 14 LOs, so only 22 combinations have data.}
Of those 22, 13 achieved good classification accuracy ($\geq$0.9) and six achieved acceptable accuracy (0.80--0.90). Three EMCS energy LOs fell below the acceptable range: Potential Energy (0.675), Conserve Energy (0.705), and Kinetic Energy (0.745). We also examine the source of this pattern in Section~\ref{sec:fit}.

% The lower classification accuracy reflects the lack of items measuring these LOs. --> solved in the discussion section.

\section{Discussion}

\color{black}

This study provides validity evidence for the student models of the MCD within the ECD framework. The student models specify the LOs the assessment targets. The evidence models supply the evidence for those LOs. Here, the Q-matrix and the DINA model serve as the evidence models, applied to items from the FCI, FMCE, and EMCS. This evidence shows whether student responses can measure and separate those LOs. We first evaluate that evidence through model fit and classification accuracy, because an item bank is useful for diagnosis if the underlying model recovers the LO structure. We then characterize the LO coverage the combined bank provides, and we situate these results relative to MCD-v1.

\subsection{Model Fit and Classification Accuracy}
\label{sec:fit}

The FCI and EMCS achieved good DINA model fit, whereas the FMCE fit only marginally (Table~\ref{tab:modfit}). The marginal FMCE fit matches a structural feature of the instrument. FMCE groups items into ``chained'' or ``blocked'' sets that share a physical scenario stem. This blocking covers 42 of the 43 scored FMCE items \citep{wells2020exploring}. The FCI blocks 13 of 30 items. The EMCS does not block items at all. Item chaining creates local item dependence among items within each block \citep{eaton2020detecting}. The DINA model assumes conditional independence between items, given a student's LO mastery profile. Local item dependence violates this assumption. Prior work also found that FMCE items engage more overlapping dimensions than FCI items \citep{laverty2018analysis}. This indicates that FMCE item design, not a modeling artifact, drives the marginal fit.

Classification accuracy met or exceeded the acceptable threshold for 19 of the 22 LO-assessment combinations. The three exceptions were EMCS energy LOs: Potential Energy, Conserve Energy, and Kinetic Energy. We examined the DINA slipping and guessing parameters and the 3PL guessing parameters for every item (Tables~\ref{tab:itempars_fci}--\ref{tab:itempars_emcs}) to investigate why. Two patterns emerged. First, EMCS items are noisier than FCI and FMCE items. The mean DINA slipping parameter is 0.349 on the EMCS versus 0.138 on the FCI and 0.136 on the FMCE. A high slip means students the model classifies as having mastered an item's required LOs still miss the item often. This weakens the evidence each response provides and lowers classification accuracy for EMCS LOs. Second, item noise alone does not explain the gap between the energy and momentum LOs on the EMCS. The momentum LOs reached acceptable-to-good accuracy (Linear Momentum 0.860, Impulse 0.820, Conserve Momentum 0.917) with slipping parameters comparable to the energy items. The difference lies in how items map onto LOs. The three energy LOs share most of their items. Eight items require all three, and any two energy LOs share about 70\% of their items (Jaccard overlap 0.67--0.73). The three momentum LOs draw on largely separate items. Any two share 18--33\% of their items, and only one item requires all three. When LOs share most of their items, few responses depend on one LO without the others, so the DINA model cannot separate mastery of one energy LO from another. This overlap reflects the conceptual nesting of the energy LOs rather than a coding error.

Kinetic energy, potential energy, and conservation of energy are not independent constructs. Conservation of mechanical energy presupposes an understanding of both kinetic and potential energy. When conceptually nested LOs share the same items, they violate the DINA model's conjunctive, independent-attribute assumption: that the LOs an item requires are distinct, and that a student must master all of them to respond correctly \citep{haertel1984application, junker2001cognitive, delaTorre2009}. This is a plausible source of the low classification accuracy for the energy LOs, distinct from the item-noise issue noted above. 

Sample size compounds this structural problem. DINA slipping and guessing estimates stabilize only with sufficiently large samples \citep{de2014cognitively, junker2001cognitive}, and the EMCS contributed the smallest sample in our data (1,990 responses), making its estimates the least stable of the three RBAs. As EMCS responses accumulate through LASSO, parameter estimates will stabilize, and classification accuracy for the energy LOs should improve even without a change in modeling approach.

Models that relax the DINA independence assumption are the next step for the structural problem. The generalized DINA (G-DINA) model \citep{de2011generalized} does not require mastery of every coded LO. It lets partial mastery of the required LOs raise the probability of a correct response, which can accommodate items that draw on several related energy ideas at once. Hierarchical diagnostic classification models \citep{templin2014hierarchical} go further. They encode prerequisite orderings among LOs, for example kinetic and potential energy as prerequisites for conservation of energy. These models are more flexible but more demanding to estimate. The DINA model estimates two parameters per item (one slip and one guess); the G-DINA model estimates up to $2^{K_j}$ parameters for an item requiring $K_j$ LOs, and needs larger samples for stable calibration \citep{de2011generalized, ma2020gdina}. A rigorous DINA--G-DINA comparison, and the estimation of hierarchical LO structures, will become feasible as energy-LO response data accumulate through LASSO. Larger samples will stabilize the noise problem, and the structural misfit still needs a more flexible model.

We considered merging Kinetic Energy, Potential Energy, and Conservation of Energy into one Energy LO. Merging would eliminate the current item overlap among these LOs and may improve classification accuracy, but it would also reduce the diagnostic resolution of the assessment. We keep them separate for three reasons. First, the three LOs match the pacing of the source texts. OpenStax covers kinetic and potential energy in separate chapters, so the energy unit spans about two weeks and four LOs (Kinetic Energy, Potential Energy, Work, and Conservation of Energy), consistent with our approximately two-LOs-per-week sizing. Second, instructors may assess the three concepts together but still need each student's mastery of each one. A merged LO would return one decision for the whole unit. Third, the overlap is a property of the current item bank rather than necessarily of the LOs themselves. The planned expansion to 35 LOs will add additional energy-specific items, which should reduce the overlap as the item bank grows. A hierarchical model treating Kinetic Energy and Potential Energy as prerequisites for Conservation of Energy could further represent this conceptual structure while preserving LO-level reporting.

Taken together, the evidence-model results provide validity evidence that the 14-LO student models are recoverable for the kinematics, forces, and momentum LOs with the current item bank and data. They also localize the weakest evidence. The energy LOs are the part of the student models that the current data and the DINA model support least well, and they are therefore the clearest target for additional data and more flexible modeling.

\subsection{The 14 Learning Objectives}

Combining the three RBAs into a single item bank increases the number of responses available per LO. DINA models with more LOs need more response data to produce stable parameter estimates — typically well over 2,000 responses \citep{ma2020gdina}. Many LOs receive items from more than one RBA. Pooling the three RBAs raises the total response evidence for those LOs beyond what any single assessment provides. 

Combining the RBAs also broadens the item bank in two ways. First, the RBAs cover different content. The FCI and EMCS split cleanly: the FCI assesses kinematics and forces, and the EMCS assesses energy and momentum, with no overlap between them. The FMCE also assesses kinematics and forces, and contributes more items than the FCI to 1D Kinematics (28 vs. 5) and Newton's Second Law (23 vs. 13), plus a small number of energy items. Second, the RBAs span different ranges of item difficulty. Combining them gives the MCD's adaptive algorithm a wider difficulty range to draw from, so it can match items to a broader span of student proficiency.
The three RBAs cover all 14 LOs with at least six items per LO, a minimum consistent with simulation studies showing that a ratio of at least four to six items per attribute supports reliable parameter estimation and classification accuracy in DINA-based analyses \citep{najera2021determining, ma2020gdina}.

The combined item bank also mixes single-LO and multi-LO items, which broadens the kinds of mastery the MCD can diagnose. Most items across all three assessments assessed one or two LOs, but the FMCE contributed the highest proportion of three-LO items, and the EMCS included a small number of items requiring four or more LOs, a complexity absent from the FCI and FMCE. This mix is an asset: single-LO items allow precise targeting of individual mastery, while multi-LO items can reveal the degree to which students integrate knowledge across related objectives \citep{cheng2009cognitive, templin2010diagnostic}. The combined item bank, therefore, offers a broader range of diagnostic complexity than any single RBA, enabling the MCD to target both isolated and integrated mastery depending on instructional need.

\subsection{Comparison with MCD-v1}

MCD-v1 \citep{le2025applying} applied the same DINA approach to the same three RBAs but organized items around four broad skills across content areas: apply vectors, conceptual relationships, algebra, and visualizations. The four-skill structure fit the FCI (RMSEA2 = 0.048, SRMSR = 0.062) and the EMCS well (RMSEA2 = 0.028, SRMSR = 0.041), and fit the FMCE only marginally (RMSEA2 = 0.090, SRMSR = 0.110). The present 14-LO structure improved fit for all three RBAs: FCI RMSEA2 fell from 0.048 to 0.033, EMCS RMSEA2 fell from 0.028 to 0.022, and FMCE RMSEA2 fell from 0.090 to 0.065. The 14-LO structure fit each RBA better than the four-skill structure did. Together, these results show that MCD-v2's LO-based student models refine MCD-v1's skill-based model by providing a more fine-grained representation of the knowledge assessed by each item. This finer structure also makes the diagnostic output more directly interpretable at the level of specific physics concepts, rather than broad cross-content skills. MCD-v2 therefore offers finer diagnostic resolution and tighter instructional alignment while preserving the psychometric viability that made the initial item bank usable.

\section{Limitations}

The DINA model requires students to master every LO an item requires. It does not represent relationships between LOs. This limits the model's ability to represent conceptually linked LOs, such as kinetic energy, potential energy, and conservation of energy. We discuss the G-DINA model and hierarchical diagnostic classification models as candidate solutions in Section~\ref{sec:fit}.

The current analysis relies on posttest data only. This limits our ability to evaluate LO mastery development during instruction. The MCD supports repeated formative use throughout a semester. Assessing its formative value requires longitudinal data across multiple points in a course. Extending the analysis to multi-timepoint data requires measurement invariance across administrations, meaning that items function the same way at each time point. Researchers must establish this invariance before interpreting any observed change in LO mastery as genuine learning. Item slipping, guessing, or LO requirements can shift between administrations. An item can then behave differently at each time point, independent of a student's true LO mastery. Testing for this instability and establishing invariance across time comes first. Otherwise, apparent LO mastery gains could reflect changes in item behavior instead of changes in student understanding.
This study focused on posttest data for a specific reason. Establishing a stable 14-LO structure at a single time point is a prerequisite for longitudinal work. Validity evidence for the student models needs to exist before evidence about change over time can be interpreted. Testing measurement invariance across pretest and posttest administrations is a direct next step. Evaluating the MCD's sensitivity to instructional gains is another. Both build on the single-time-point validity evidence established here.

The analysis does not examine differential item functioning (DIF) across demographic groups (e.g., gender, race, course type). Prior work has documented DIF on several FCI items and on one FMCE item \citep{traxler2018gender, henderson2018item}. Later work found the five-factor model of the FCI measurement invariant across the intersections of race and gender \citep{morley2023measurement}. A DIF analysis across those intersections found large DIF on many FCI items, with patterns that followed the same five-factor structure, and concluded that the items likely measure real differences in physics knowledge across groups rather than item bias \citep{buncher2025force}. We have not run a comparable analysis for MCD-v2. Testing its LO-level DIF across demographic groups is a next step and will show if its diagnostic feedback is fair across student populations.
 
This study leaves several practical questions untested: whether instructors can interpret LO-level feedback, whether that feedback changes teaching decisions, and whether students who receive it learn more than students who do not. Several features of the MCD's design support the case that it will prove useful. The LOs align with the weekly pacing structure common in mechanics courses, so LO-level feedback maps onto decisions instructors already make about what to cover next. The item bank draws from RBAs that physics courses already administer, so the MCD adds diagnostic resolution without requiring instructors to adopt an unfamiliar assessment. The LASSO platform already reaches hundreds of courses and institutions, which lowers the barrier to piloting the MCD in real classrooms. Testing these claims directly is a necessary next step: whether instructors find LO-level feedback actionable, and whether using it changes student outcomes.

\section{Conclusions}

This study shows that CD analysis can extract LO-level information from the same items instructors already use. Combining three common RBAs into an MCD item bank yields LO-level inferences that no single instrument can provide on its own. A CD model, applied to existing RBA items, turns fixed-length, single-score assessments into a source of fine-grained diagnostic information without requiring new items. CD-CATs, such as the MCD, extend this CD foundation by selecting items that maximize diagnostic information at each administration step, increasing both precision and efficiency. By providing this framework and tool, we aim to better support physics instructors and students throughout introductory mechanics courses.

The 14-LO structure performed well for kinematics, forces, and momentum LOs. Two limitations remain, marginal FMCE fit and lower classification accuracy for three EMCS energy LOs. Section~\ref{sec:fit} traces these to specific, plausible sources. FMCE's marginal fit likely reflects its design. Most FMCE items come in sets that share one scenario, so responses within a set depend on one another, and the DINA model assumes independent items. The three energy LOs' lower classification accuracy has three likely sources. EMCS items have higher slipping parameters than FCI and FMCE items, the three energy LOs share most of their items, and the EMCS contributed the smallest sample (1,990 responses). Combining the three RBAs offers a partial remedy, but an uneven one. We fit the DINA model separately for each RBA, so FMCE's fit statistic reflects only its own response data. Combining RBAs helps the energy LOs more. Kinetic Energy and Potential Energy each draw items from both the FMCE and EMCS, giving each more pooled response evidence than either RBA provides alone. Conserve Energy, in contrast, draws items only from the EMCS, so combining RBAs cannot yet supplement its evidence. The item overlap may also mean the current items cannot separate the three energy LOs. We keep them separate because they match the pacing of the source texts and because instructors need mastery decisions for each one. The planned expansion to 35 LOs will add energy-specific items, while a hierarchical model can further test whether Kinetic Energy and Potential Energy function as prerequisites for Conservation of Energy.

Our findings align with those of MCD-v1 \citep{le2025applying}, which organized items from the same three RBAs around four broad cognitive skills rather than content-specific LOs. Fit improved under the present LO-based structure for all three RBAs, and the FMCE remained the only marginal-fit assessment under both structures — a pattern that recurs across two independently coded attribute structures and likely reflects a property of the FMCE items themselves rather than either coding scheme. This progression suggests MCD-v2's LO-based student models refine MCD-v1's skill-based model. MCD-v2 offers finer diagnostic resolution and closer instructional alignment while preserving the psychometric viability of the item bank.

Future work will support instructors and researchers with timely, actionable, fine-grained diagnostic feedback throughout introductory mechanics instruction. Three priorities will drive this work.
First, the current item bank does not yet cover all 35 planned LOs. The FCI, FMCE, and EMCS do not include items for rotational mechanics or mathematical reasoning, topics outside these assessments' conceptual focus. Adding existing RBAs built for these topics can close these gaps: the Rotational and Rolling Motion Conceptual Survey (RRMCS) \citep{rimoldini2005student} for rotational mechanics, and the Test of Understanding Graphs in Kinematics (TUG-K) \citep{beichner2011test} for mathematical reasoning and graphical interpretation.
Second, the MCD needs a way to add new items without pausing testing to try them out separately first. Online calibration solves this: the system embeds new, unscored items alongside the operational item bank, collects real student responses to them during normal use, and estimates their DINA parameters from those responses \citep{wang2021adaptive, yu2024iterative}. Once an item's parameters are stable, the system can add it to the scored item bank. This lets the item bank grow while the MCD stays in use, without a separate pretesting phase.
Third, as more courses use the LASSO platform, which is free for instructors, response data will continue to accumulate across all three RBAs. Growth in EMCS responses in particular will let future work re-estimate DINA parameters for the energy LOs with a larger sample than the one used here, and compare the DINA model directly against G-DINA and hierarchical alternatives, as discussed in Section~\ref{sec:fit}.

\color{black}

\section*{Acknowledgments}
This work was funded through NSF awards \#2141847 and \#2526720. The data analyzed in this study were collected through the LASSO platform and are available to researchers upon request through the LASSO data sharing process at \textcolor{blue}{https://lassoeducation.org.}

\section*{Appendix}

\begin{table*}[t]
\centering
\caption{\justifying Learning objectives and their corresponding items from the FCI, FMCE, and EMCS. 
Note: Each row represents one LO, and each column lists the item numbers from the corresponding assessment that map to that LO. Empty cells indicate that the assessment does not include items for that LO.}
\label{tab:topics}
\setlength{\tabcolsep}{5pt} 
\begin{tabularx}{\textwidth}{
  >{\raggedright\arraybackslash}p{4.2cm}
  >{\raggedright\arraybackslash}X
  >{\raggedright\arraybackslash}X
  >{\raggedright\arraybackslash}X
}
\hline
\hline
Learning Objective &  FCI &  FMCE &  EMCS \\
\hline

  Vectors & 6, 7, 8, 9, 12, 13, 14, 21, 22, 23 & 8, 9, 10, 27, 28, 29 &  \\
  \hline
  1D Kinematics & 19, 20, 24, 26, 27 & 1, 2, 3, 4, 5, 6, 7, 11, 12, 13, 14, 16, 17, 18, 19, 21, 22, 23, 24, 25, 26, 27, 28, 29, 40, 41, 42, 43 & \\ \hline
  Free Fall & 1, 3, 13 & 11, 12, 13, 27, 28, 29 &  \\  
  \hline
  2D Kinematics & 2, 9, 12, 14, 21, 22 & &  \\ 
  \hline
  Free-Body Diagrams & 5, 11, 17, 18, 25, 26, 27, 30 & 1, 2, 3, 4, 5, 6, 7, 8, 9, 10 &  \\ 
  \hline
  Newton’s Second Law & 1, 3, 6, 7, 8, 10, 17, 22, 23, 24, 25, 26, 27 & 1, 2, 3, 4, 5, 6, 7, 8, 9, 10, 14, 15, 16, 17, 18, 19, 20, 21, 22, 23, 24, 25, 26 &  \\ 
  \hline
  Newton’s Third Law &  4, 5, 15, 16, 28  & 30, 31, 32, 33, 34, 35, 36, 37, 38, 39 & \\
  \hline
  Kinetic Energy & & 44, 46 & 2, 4, 13, 14, 15, 16, 17, 20, 22, 24 \\
  \hline
  Potential Energy & & 44, 45, 46, 47 & 1, 2, 4, 8, 13, 14, 15, 16, 20, 22 \\
  \hline
  Work & & & 1, 6, 8, 9, 12, 20, 24, 25 \\
  \hline
  % Work-Energy Theorem  & & & 1, 8, 9, 24, 25 \\
  % \hline
  Conserve Energy & & & 2, 4, 9, 13, 14, 15, 16, 20, 22 \\
  \hline
  Linear Momentum & & & 7, 10, 13, 14, 18, 23 \\
  \hline
  Impulse & & & 3, 14, 16, 18, 19, 23 \\
  \hline
Conserve Momentum & 
& 
& 
3, 5, 10, 11, 14, 16, 21 \\
\hline
\hline
\end{tabularx}
\end{table*}

\begin{table}[t]
\centering
\caption{\justifying FCI item parameters: DINA guessing ($g$) and slipping ($s$); 3PL discrimination ($a$), difficulty ($b$), and pseudo-guessing ($c$). Item 29 is omitted; we excluded it from all analyses (Section~\ref{sec:data}).}
\label{tab:itempars_fci}
\setlength{\tabcolsep}{8pt}
\begin{tabular}{c c c c c c}
\hline
\hline
Item & $g$ & $s$ & $a$ & $b$ & $c$ \\
\hline
1  & 0.691 & 0.049 & 1.56 & -0.51 & 0.473 \\
2  & 0.394 & 0.176 & 2.32 & 0.53  & 0.363 \\
3  & 0.424 & 0.120 & 1.61 & -0.17 & 0.183 \\
4  & 0.352 & 0.061 & 1.39 & -0.92 & 0.002 \\
5  & 0.208 & 0.215 & 2.72 & 0.44  & 0.147 \\
6  & 0.716 & 0.022 & 1.83 & -0.82 & 0.426 \\
7  & 0.582 & 0.061 & 1.79 & -0.37 & 0.376 \\
8  & 0.482 & 0.091 & 1.95 & -0.04 & 0.358 \\
9  & 0.266 & 0.213 & 2.31 & 0.53  & 0.190 \\
10 & 0.454 & 0.049 & 2.07 & -0.56 & 0.158 \\
11 & 0.225 & 0.165 & 1.89 & 0.08  & 0.050 \\
12 & 0.692 & 0.035 & 1.62 & -0.65 & 0.372 \\
13 & 0.166 & 0.067 & 3.11 & 0.01  & 0.066 \\
14 & 0.300 & 0.178 & 1.98 & 0.41  & 0.189 \\
15 & 0.230 & 0.269 & 0.87 & -0.23 & 0.004 \\
16 & 0.552 & 0.064 & 1.51 & -1.13 & 0.092 \\
17 & 0.239 & 0.255 & 2.11 & 0.56  & 0.159 \\
18 & 0.244 & 0.101 & 2.95 & 0.12  & 0.152 \\
19 & 0.261 & 0.111 & 1.87 & -0.18 & 0.149 \\
20 & 0.219 & 0.165 & 1.39 & -0.27 & 0.007 \\
21 & 0.158 & 0.361 & 2.13 & 0.88  & 0.123 \\
22 & 0.338 & 0.113 & 3.22 & 0.43  & 0.286 \\
23 & 0.249 & 0.151 & 2.47 & 0.21  & 0.182 \\
24 & 0.526 & 0.038 & 1.99 & -0.54 & 0.218 \\
25 & 0.225 & 0.201 & 3.24 & 0.53  & 0.191 \\
26 & 0.084 & 0.302 & 2.89 & 0.66  & 0.037 \\
27 & 0.456 & 0.123 & 1.90 & 0.13  & 0.280 \\
28 & 0.194 & 0.039 & 1.97 & -0.60 & 0.002 \\
30 & 0.190 & 0.214 & 2.36 & 0.34  & 0.112 \\
\hline
\hline
\end{tabular}
\end{table}

\begin{table*}[t]
\centering
\caption{\justifying FMCE item parameters: DINA guessing ($g$) and slipping ($s$); 3PL discrimination ($a$), difficulty ($b$), and pseudo-guessing ($c$).}
\label{tab:itempars_fmce}
\setlength{\tabcolsep}{5pt}
\begin{tabular*}{\textwidth}{@{\extracolsep{\fill}} c c c c c c c c c c c c}
\hline
\hline
Item & $g$ & $s$ & $a$ & $b$ & $c$ & Item & $g$ & $s$ & $a$ & $b$ & $c$ \\
\hline
1 & 0.227 & 0.066 & 7.04 & 0.73 & 0.222 & 25 & 0.142 & 0.188 & 3.99 & 0.58 & 0.082 \\
2 & 0.093 & 0.177 & 4.67 & 0.80 & 0.071 & 26 & 0.264 & 0.102 & 3.22 & 0.29 & 0.103 \\
3 & 0.242 & 0.070 & 5.31 & 0.65 & 0.199 & 27 & 0.209 & 0.016 & 4.24 & 0.48 & 0.112 \\
4 & 0.170 & 0.103 & 6.05 & 0.74 & 0.154 & 28 & 0.148 & 0.042 & 4.11 & 0.57 & 0.078 \\
5 & 0.255 & 0.102 & 3.78 & 0.59 & 0.170 & 29 & 0.282 & 0.010 & 4.10 & 0.43 & 0.172 \\
6 & 0.071 & 0.442 & 3.91 & 1.11 & 0.061 & 30 & 0.263 & 0.036 & 1.46 & -0.13 & 0.043 \\
7 & 0.254 & 0.104 & 3.99 & 0.62 & 0.179 & 31 & 0.215 & 0.035 & 1.88 & 0.06 & 0.077 \\
8 & 0.095 & 0.184 & 5.86 & 0.91 & 0.063 & 32 & 0.182 & 0.011 & 2.18 & 0.11 & 0.088 \\
9 & 0.100 & 0.172 & 5.65 & 0.89 & 0.064 & 33 & 0.704 & 0.006 & 1.86 & -1.30 & 0.000 \\
10 & 0.280 & 0.065 & 5.84 & 0.75 & 0.222 & 34 & 0.166 & 0.016 & 2.30 & 0.17 & 0.100 \\
11 & 0.076 & 0.075 & 5.62 & 0.65 & 0.095 & 35 & 0.339 & 0.165 & 2.03 & 0.27 & 0.208 \\
12 & 0.085 & 0.081 & 5.50 & 0.63 & 0.094 & 36 & 0.118 & 0.406 & 1.87 & 0.97 & 0.106 \\
13 & 0.205 & 0.023 & 5.88 & 0.55 & 0.201 & 37 & 0.415 & 0.158 & 1.62 & -0.11 & 0.124 \\
14 & 0.063 & 0.239 & 4.83 & 0.72 & 0.055 & 38 & 0.092 & 0.425 & 2.07 & 1.01 & 0.099 \\
15 & 0.780 & 0.039 & 1.41 & -1.59 & 0.001 & 39 & 0.280 & 0.091 & 2.14 & 0.03 & 0.089 \\
16 & 0.101 & 0.170 & 6.28 & 0.67 & 0.103 & 40 & 0.287 & 0.067 & 1.52 & -0.97 & 0.000 \\
17 & 0.045 & 0.399 & 3.35 & 0.95 & 0.043 & 41 & 0.190 & 0.285 & 1.31 & -0.21 & 0.005 \\
18 & 0.048 & 0.264 & 6.43 & 0.75 & 0.052 & 42 & 0.124 & 0.120 & 1.66 & -0.55 & 0.001 \\
19 & 0.072 & 0.275 & 5.70 & 0.77 & 0.076 & 43 & 0.485 & 0.057 & 1.46 & -1.34 & 0.000 \\
20 & 0.104 & 0.382 & 4.79 & 0.89 & 0.097 & 44 & 0.295 & 0.013 & 2.37 & 0.48 & 0.245 \\
21 & 0.085 & 0.342 & 4.41 & 0.79 & 0.057 & 45 & 0.361 & 0.003 & 2.19 & 0.34 & 0.278 \\
22 & 0.299 & 0.085 & 3.38 & 0.30 & 0.156 & 46 & 0.257 & 0.020 & 2.35 & 0.64 & 0.255 \\
23 & 0.174 & 0.137 & 4.07 & 0.48 & 0.091 & 47 & 0.258 & 0.023 & 2.39 & 0.59 & 0.249 \\
24 & 0.238 & 0.110 & 3.49 & 0.36 & 0.109 &    &       &       &      &      &       \\
\hline
\hline
\end{tabular*}
\end{table*}

\begin{table}[t]
\centering
\caption{\justifying EMCS item parameters: DINA guessing ($g$) and slipping ($s$); 3PL discrimination ($a$), difficulty ($b$), and pseudo-guessing ($c$).}
\label{tab:itempars_emcs}
\setlength{\tabcolsep}{8pt}
\begin{tabular}{c c c c c c}
\hline
\hline
Item & $g$ & $s$ & $a$ & $b$ & $c$ \\
\hline
1  & 0.319 & 0.220 & 1.09 & 0.27  & 0.008 \\
2  & 0.504 & 0.099 & 1.11 & -0.65 & 0.007 \\
3  & 0.373 & 0.352 & 0.62 & 0.70  & 0.073 \\
4  & 0.358 & 0.190 & 2.41 & 0.76  & 0.339 \\
5  & 0.170 & 0.283 & 2.63 & 0.95  & 0.143 \\
6  & 0.197 & 0.431 & 1.95 & 1.28  & 0.169 \\
7  & 0.405 & 0.072 & 1.51 & -0.50 & 0.003 \\
8  & 0.168 & 0.425 & 1.57 & 1.29  & 0.109 \\
9  & 0.210 & 0.560 & 1.07 & 1.91  & 0.142 \\
10 & 0.194 & 0.316 & 2.00 & 1.06  & 0.145 \\
11 & 0.249 & 0.212 & 1.75 & 0.58  & 0.098 \\
12 & 0.201 & 0.547 & 1.97 & 1.76  & 0.202 \\
13 & 0.329 & 0.384 & 0.83 & 1.14  & 0.164 \\
14 & 0.320 & 0.382 & 1.63 & 1.47  & 0.276 \\
15 & 0.268 & 0.229 & 1.25 & 0.26  & 0.013 \\
16 & 0.465 & 0.416 & 1.50 & 2.37  & 0.460 \\
17 & 0.240 & 0.500 & 1.40 & 1.86  & 0.300 \\
18 & 0.334 & 0.269 & 0.83 & 0.22  & 0.028 \\
19 & 0.218 & 0.246 & 1.11 & 0.47  & 0.113 \\
20 & 0.289 & 0.188 & 1.27 & 0.35  & 0.008 \\
21 & 0.344 & 0.327 & 2.14 & 1.28  & 0.331 \\
22 & 0.214 & 0.502 & 1.44 & 1.58  & 0.187 \\
23 & 0.326 & 0.589 & 3.34 & 1.82  & 0.319 \\
24 & 0.115 & 0.572 & 2.20 & 1.63  & 0.114 \\
25 & 0.211 & 0.421 & 0.97 & 1.05  & 0.037 \\
\hline
\hline
\end{tabular}
\end{table}

\bibliography{ref.bib}
\end{document}